\documentclass[prl,twocolumn]{revtex4}
\usepackage{graphicx,amsmath,amssymb}
\renewcommand{\narrowtext}{\begin{multicols}{2} \global\columnwidth20.5pc}

\def\be{\begin{eqnarray}}
\def\ee{\end{eqnarray}}

\newcommand{\Eq}[1]{Eq.~(\ref{#1})}

\begin{document}
\draft

\title{Abelian and Non-abelian Hall Liquids and Charge Density Wave: Quantum Number
Fractionalization in One and Two Dimensions}

\author{Alexander Seidel and Dung-Hai Lee}

\affiliation{Department of Physics,University of California at
Berkeley, Berkeley, CA 94720, USA}
 \affiliation{
Material Science Division, Lawrence Berkeley National
Laboratory,Berkeley, CA 94720, USA.}

\date{\today}

\begin{abstract}

Previously we have demonstrated that, on a torus, the abelian
quantum hall liquid is adiabatically connected to a charge density
wave as the smaller dimension of the torus is varied.\cite{seidel}
In this work we extend this result to the non-abelian bosonic Hall state.
The outcome of these works is the realization that the
paradigms of quantum number fractionalization in one dimension
(polyacetylene)\cite{su} and two dimensions (fractional quantum
Hall effect)\cite{laughlin} are in fact equivalent.

\end{abstract}

\maketitle

Quantum number fractionalization is a subject of considerable
interest in condensed matter physics  recently. Two well-known
examples where such phenomenon occurs are (1) the soliton in a
one-dimensional (1D) charge density wave\cite{su}, and (2) the
quasiparticles in the fractional quantum Hall
effect\cite{laughlin}. Usually they are regarded as distinct
mechanisms of charge fractionalization.

In a recent paper, Seidel {\it et al} have shown that when placed
on a $L_1\times\infty$ torus, the $\nu=1/3$ fractional quantum
Hall liquid is adiabatically connected to a period-three charge
density wave when $L_1$ is varied\cite{seidel}. In a subsequent
work the same conclusion was reached independently\cite{karl2}.
Such continuity is very powerful in identifying discrete quantum
numbers, since the latter cannot change during an adiabatic
process. For example, the topological three fold-degeneracy in the
large $L_1/l_B$ limit ($l_B$ is the magnetic length) is
adiabatically related to the three charge density wave patterns at
small $L_1/l_B$. Moreover, the charge 1/3 quasiparticles at large
$L_1/l_B$ are
 evolved from the Su-Schrieffer domain walls at small
$L_1/l_B$\cite{seidel}.

The above results are based on a 2D $\rightarrow$ 1D mapping
discussed in Ref.\cite{ll,karl}. Specifically, in the Landau
gauge, and after projection to the lowest Landau level, the
pseudopotential Hamiltonian\cite{tk}, for which the Laughlin state
is the exact ground state, becomes a center-of-mass position
conserving pair hopping model on a 1D lattice\cite{ll,seidel}. The
torus parameter $L_1/l_B$ becomes the hopping range. Within this
model it can be shown that for all non-zero $L_1/l_B$ there is an
energy gap separating the three degenerate ground states from the
excited states. In addition, each ground state has a
different center-of-mass position, and exhibits a non-zero charge
density wave order as long as $L_1/l_B$ is finite\cite{hr,seidel}.
Finally, due to the simultaneous conservation of the center of
mass position and momentum, all eigenstates are at least
three-fold degenerate.

In this paper, we apply the same mapping to study the bosonic
non-abelian quantum Hall state at filling factor
$\nu=1$\cite{mooreread}. Our motivations are as follows. First,
there is a general argument due to Oshikawa\cite{oshikawa}, which
says that at filling factor $\nu=p/q$, if a system possesses an
excitation gap, then the ground state must be at least q-fold
degenerate on a (d-dimensional) torus. 
From this point of view, one expects that in the absence of hidden
symmetry, the ground state for a gapped $\nu=1$ system should be
non-degenerate. However, based on analyzing the ground state
wavefunction, it was shown that the gapped non-Abelian bosonic
Hall state is three-fold degenerate on torus\cite{greiter,readrezayi}.
Second, we are curious whether the adiabatic connection in the
$\nu=1/3$ abelian quantum Hall state
also holds true for the non-abelian case. 

In the infinite 2D plane the bosonic Pfaffian state\cite{mooreread}
\be\label{pfaff}
  \Psi=Pf\Big[{1\over z_i-z_j}\Big]\prod_{(ij)} (z_i-z_j)~
  \exp[-\sum_k|z_k|^2/4].
\ee is the unique  ground state wavefunction of the following
three-particle pseudopotential Hamiltonian at $\nu=1$
\begin{equation}\label{V}
  H= \sum_{(ijk)}\delta(z_i-z_j)\delta(z_i-z_k).
\end{equation}
When the Pfaffian state is generalized to toroidal
geometry\cite{greiter,readrezayi}, a threefold degeneracy emerges. As
emphasized in the beginning, unlike the Abelian quantum Hall
state, this degeneracy cannot be attributed to different
center-of-mass quantum numbers.

Following our procedure in Ref.\cite{seidel} we first establish a
simple picture for this degeneracy by making the smaller dimension
of the torus much smaller than the magnetic length $l_B$. To this
end, we express the Hamiltonian of \Eq{V} in the Landau gauge
basis of lowest Landau level of a $L_1\times L_2$ torus and obtain
$H=\sum_R Q_R^\dagger Q_R$, where \be\label{H}
Q_R=\sum_{\substack{m+n+p=3R\,\text{mod}\,N}}f(R-m,R-n,R-p)~c_m
c_n c_p\ee  and $f(a,b,c)=\kappa \sum_{s,t}
\exp[-((a+sN)^2+(b+tN)^2+(c-(s+t)N)^2)/2].$
 In the above $m,n,p$
runs over the index of the lowest Landau level orbitals. One may
identify each orbital with a site on a 1D ring. The
total number of lattice site, $N$, is equal to the number of
magnetic flux quanta piercing through the surface of the
torus,i.e., $N=L_1L_2/2\pi l_B^2$. The operator $c_n$ annihilates
a boson on the $n$-th site. \Eq{H} describes center-of-mass
conserving triplet hopping. The only parameter, $\kappa$, of this
Hamiltonian, $\kappa=2\pi l_B/L_1$, sets the hopping range.

{\bf Symmetry~} The Hamiltonian in \Eq{H} not only commutes with the
lattice translation operator, $T$,  but also commute with the
operator \be
  U=\exp[2\pi i/N \sum_n n c^\dagger_n c_n] .
\ee  We identify $U$ as the center-of-mass position operator of
the ring. For filling factor $\nu=p/q$ it is simple to prove that
$U T=e^{i 2\pi p/q} T U.$ Thus for $\nu=1$ the two operators $T$
and $U$ commute. As the result we can label any energy eigen state
by three quantum number, i.e., the eigenvalue with respect to
$H,U$ and $T$, i.e $|E,u,t\rangle$ where $H|E,u,t\rangle = E
|E,u,t\rangle,~ U|E,u,t\rangle=\exp(iu)|E,u,t\rangle$, and
$T|E,u,t\rangle=\exp(it)|E,u,t\rangle.$

In addition to the above symmetries, \Eq{H} has the remarkable
property that it is  ``self-dual'' under Fourier
transformation\cite{seidel}. That is, when the model is
reexpressed in momentum space, which is a lattice of discrete
momenta, $2\pi n/N$, one obtains a Hamiltonian which is identical
to the real space one, except for an inversion of the parameter
$\kappa\rightarrow \tilde{\kappa}=2\pi/\kappa N$. Thus a
Hamiltonian at $\kappa$ in the real space representation is
identical to the Hamiltonian at $\tilde\kappa$ in momentum space.
Consequently, the spectrum of \Eq{H} at $\kappa$ is identical to
that  at $\tilde\kappa =2\pi/\kappa N$. Since the quantum number
$u$ and $t$ interchanges upon going from the real  to the momentum
space and vice versa, it follows that all states at $\kappa$ that
have quatum number $Q=(u,t)$ are degenerate with states at
$\tilde\kappa$ with quantum number $Q=(t,u)$. 
\begin{figure}
\includegraphics[width=6cm]{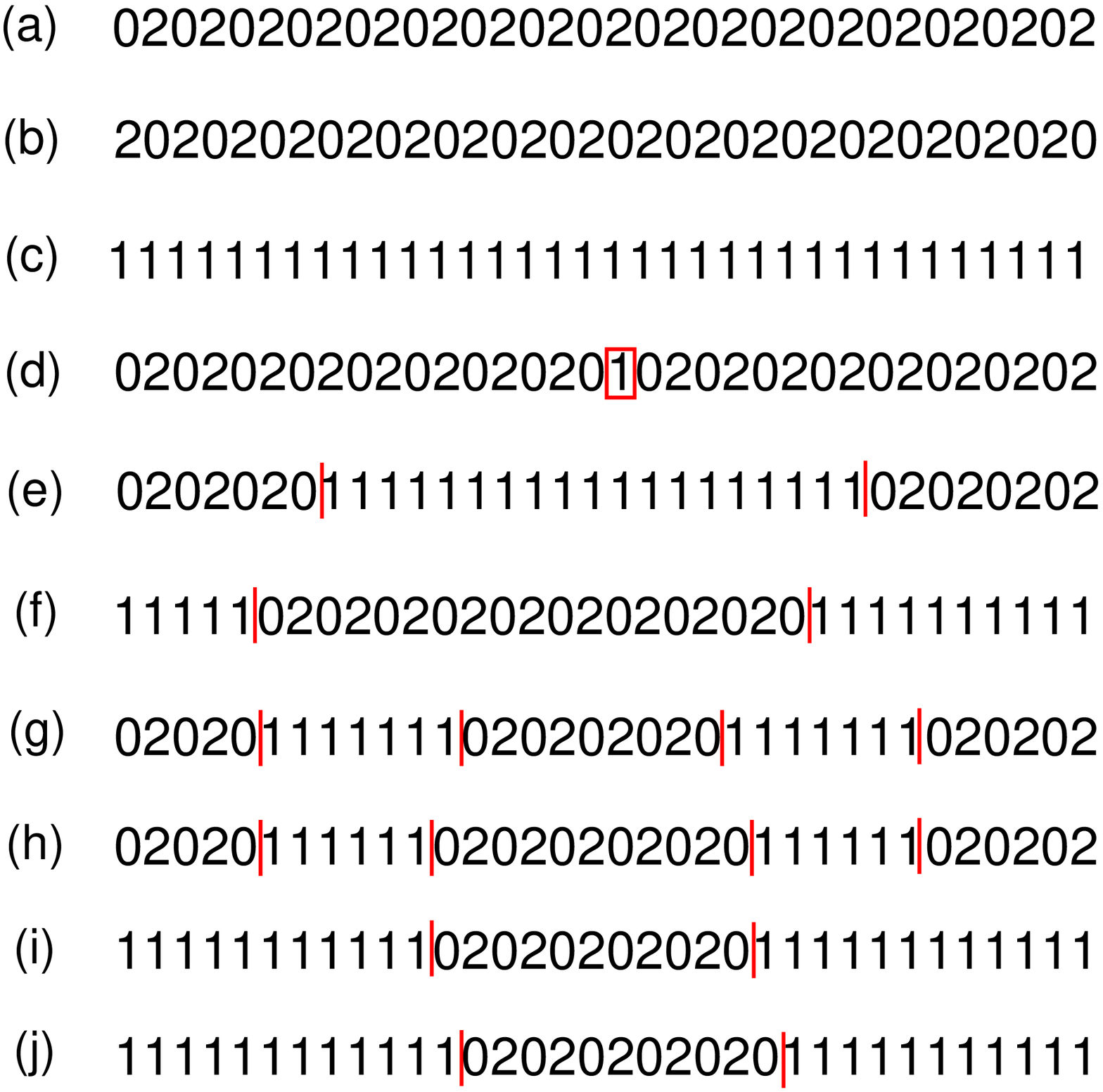}
\caption { The large $\kappa$ limits of the ground state (a)-(c),
2-hole excited states (d)-(f) and (i),(j), and 4-hole excited
state (g),(h). A  vertical segment  indicated the position of a charge
$\Delta Q=1/2$ domain wall, see text. }\label{fig1}
\end{figure}

{\bf The large $\kappa$ limit~} Since for any $L_1/l_B$, the three
Pfaffian ground states\cite{greiter,readrezayi} are zero energy eigenstates
of \Eq{V}, it follows that for any $\kappa$ there must be three
ground states each annihilated by all the operators $Q_R$. We will
now solve \Eq{H} in the limit $\kappa\gg 1$. By keeping the two
largest hopping amplitudes we obtain
\begin{equation}
  \label{largek}
H\approx\kappa^2\sum_n[(c_n^\dagger)^3 (c_n)^3+
\exp(-\kappa^2/3)(c_n^\dagger)^2 c_{n\pm 1}^\dagger (c_n)^2
c_{n\pm 1}]
\end{equation}
Such a Hamiltonian penalizes states that have three particles
occupying the same site, or three particles occupying two adjacent
sites. All other terms in the Hamiltonian are exponentially
smaller than the ones retained. The three zero energy states of
\Eq{largek} are shown in Fig.1(a)-(c). Among them the two states
in panel (a) and (b) break the translational symmetry. The third
state shown in panel (c) is translationally {\em invariant}. These
states will be referred to as $(02)$-, $(20)$- and $(11)$-state
from now on. Thus, as in the abelian-case\cite{seidel}, we do
obtain a simple pictures for the ground states in the large
$\kappa$ limit. The primary question to be answered now is whether
the low energy physics at large $\kappa$ is in fact {\em
adiabatically connected} to the limit $\kappa\rightarrow 0$, as we
argued in the abelian case. 

{\bf The degeneracy} Before we turn to this central issue, we
first highlight some crucial differences between the abelian and
the non-abelian problem. First, while in both cases the correct
ground state degeneracy becomes manifest in the large-$\kappa$
limit, for the abelian case this degeneracy is dictated by (i)
translation invariance, (ii) conservation of the center-of-mass
position, and (iii) the non-commutivity between $T$ and $U$. Thus
it holds true at any value of $\kappa$. 
Theis 
argument is
no longer valid in the present case, because $T$ and $U$ commute.
Furthermore, {\em all} of the ground state in the abelian case are
related by a simple lattice translation, and are thus 
energetically equivalent. Theis is not true in the present case,
since at large $\kappa$ only two of the ground states are related
by translation. Indeed, the $(11)$-state appears as special when
compared to the $(02)$- and $(20)$ states, and it seems that a
nearest neighbor density-density interaction could lift the
threefold degeneracy. While this is true at large $\kappa$, we
believe that such lifting of degeneracy will become exponentially
small as $\kappa\rightarrow 0$,  and the threefold degeneracy will
become robust at small $\kappa$. This is because, as we shall show
later, the order parameter associated with the (02) and (20) state
exponentially vanishes as $\kappa\rightarrow 0$.

We now show that assuming the existence of adiabatic continuity
(to be shown below),
the 3-fold ground state degeneracy at any $\kappa$ can be
explained as a consequence of the existence of the 2-fold
degenerate symmetry broken states at large $\kappa$ {\em alone}.
At large $\kappa$ we may form symmetric and antisymmetric linear
combination of the two symmetry breaking ground states, so that
they become eigenstates of $T$ with eigenvalues $t=0$ and $t=\pi$.
It is simple to show that both (02) and (20) have eigenvalue $u=0$
with respect to the center-of-mass operator $U$. Consequently the
newly formed linear combinations also have the same eigenvalue.
Thus at large $\kappa$ we have a degenerate pair of ground states
with quantum numbers $Q=(0,0)$ and $Q=(0,\pi)$. If adiabatic
continuity exists, we must have ground states with these quantum
numbers at any $\kappa$, including $\tilde{\kappa}$. Now let us
consider applying the duality transformation to the $Q=(0,\pi)$
ground state at $\tilde{\kappa}$. As we explained earlier, this
transforms it to a state $Q=(\pi,0)$ at $\kappa$. Since the ground
state energy at $\kappa$ and $\tilde{\kappa}$ is the same, this
transformed state must have the same energy as the $Q=(0,0)$ and
$Q=(0,\pi)$ ground states at $\kappa$. Thus we have generated a
third orthogonal state degenerate with the $Q=(0,0),(0,\pi)$
ground state at the same value of $\kappa$. Consequently the
ground state is at least three-fold degenerate. In the large
$\kappa$ limit the $Q=(\pi,0)$ state is the $(11)$ ground state.

\begin{figure}
\includegraphics[width=6cm]{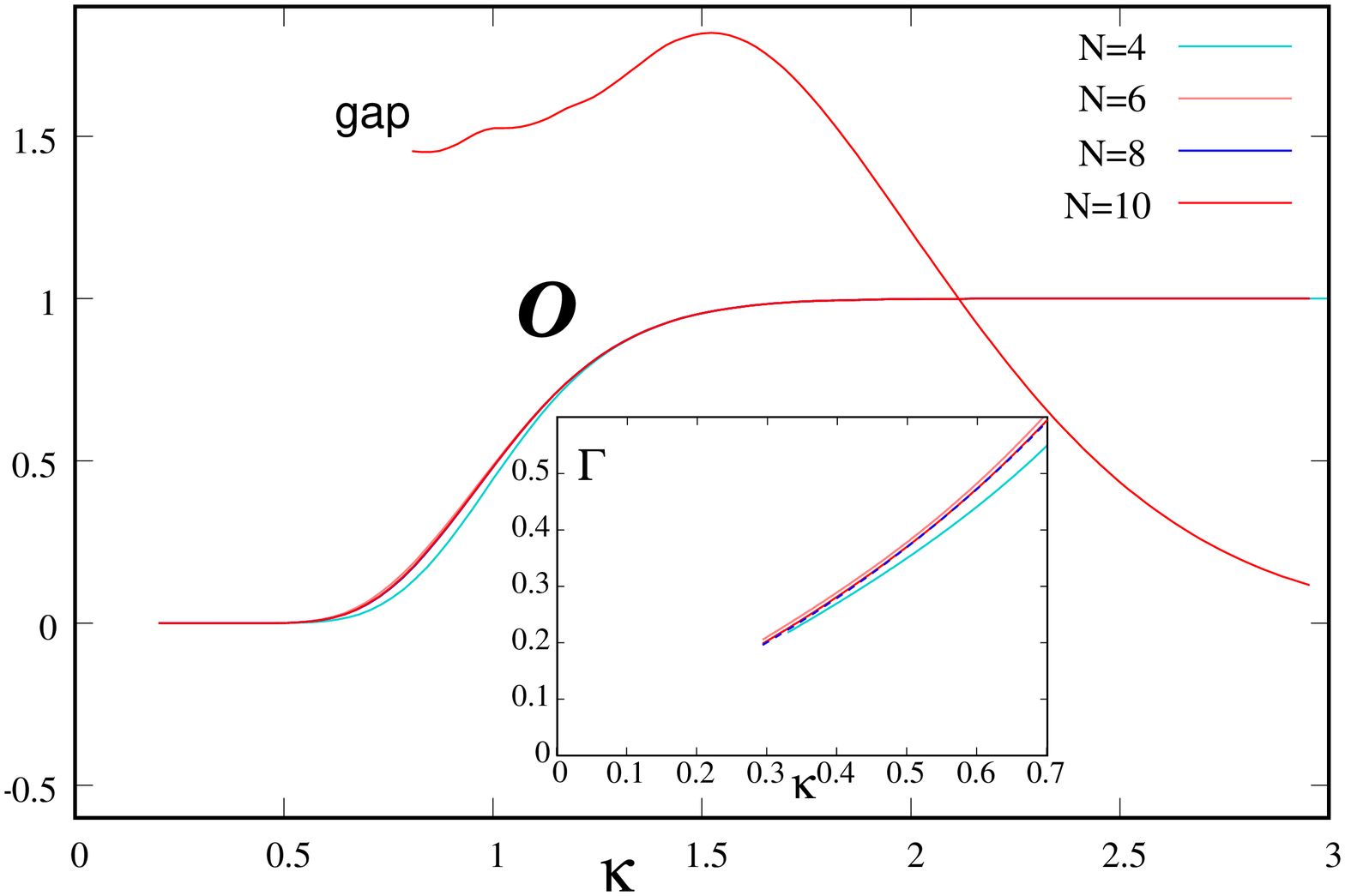}
\caption {Colored curves: the charge density wave order parameter
$\cal O$ of the ground states in the zero center-of-mass sector
for $N=4,6,8,10$ particles. These states are adiabatically
connected to the $(20)$- and $(02)$-states in Fig 1(a) and (b).
The ground state that is connected to the $(11)$ state of Fig.
1(c) is free of symmetry breaking, hence has ${\cal O}=0$. The
single curve labeled ``gap'' shows the energy gap for $N=10$  
versus $\kappa$ for
$\kappa>\kappa_{sd}$. Gap data is scaled down by a factor of $4$
for convenience. Inset: The quantity $\Gamma=1/\sqrt{-\log{\cal
O}}$. }\label{data}
\end{figure}

{\bf Adiabatic continuity} Fig 2) displays the charge density wave
order parameter ${\cal O}=\sum_n (-1)^n c_n^\dagger c_n/N$
associated with the two symmetry breaking ground states in the
$u=0$ sector of \Eq{H}. As explained above, these states are
evolved from the $(20)$ and $(02)$ states at large $\kappa$. This
is manifested by the fact that $|{\cal O}|$ becomes $1$,
reflecting perfect order, in the large $\kappa$ limit. We note
that the value of ${\cal O}$ converges remarkably fast with
increasing system size (Fig. 2), and it seems hardly necessary to
look at larger systems for this purpose. The order parameter
seemingly vanishes around $\kappa\approx .5$, very 
similar to the
behavior observed in the abelian case\cite{seidel}. This could
lead one to think that there is a phase transition near this
point, below which the order parameter is zero in the
$N\rightarrow\infty$  limit. However, a careful study of the order
parameter at small $\kappa$ suggests otherwise. In the inset we
plot the quantity $\Gamma$ which is related to the order parameter
via ${\cal O}=\exp(-1/ \Gamma^2)$. Although numerical precision
limits us to $\kappa\gtrsim .29$, we observe that the order
parameter continues to display a rapidly converging, alternating
behavior as a function of system size below $\kappa=.5$. The fact
that $\Gamma$ converges to nonzero values as a function of particle
number for all $\kappa$ value we studied clearly implies that for
these $\kappa$ value the limiting value ($N\rightarrow\infty$)  of
the order parameter is nonzero. Moreover, it seems likely that the
limiting $\Gamma$-curve extrapolates into the origin. This implies
that as $N\rightarrow\infty$ the order parameters associated with
the evolved (02) and (20) states stay non-zero except at $\kappa=0$
(infinite $L_1$). 
This is fully 
consistent with our findings in the abelian
case\cite{hr,seidel}. Thus the analysis based on the order
parameter suggests that there is no phase transition. A new
feature that occurs in the non-abelian case is the existence of a
non-symmetry-breaking ground state, i.e, the descendent of the
(11) state, at all values of $\kappa$. Due to the exponentially
small order parameter in Fig.2, the distinction between the two
kinds of ground states becomes academic as $\kappa$ decreases
below $0.5$.

Now we demonstrate that the exponentially small order parameter at
small $\kappa$ does not affect the robustness of the energy gap.
Fig. 2 shows the energy gap for $10$ particles as a function of
$\kappa$. The data is truncated below the self-dual point
$\kappa_{sd}=\sqrt{2\pi/N}$ because this part of the data is just
the dual image of that above  $\kappa_{sd}$. In addition, due to
the  strong size dependence of $\kappa_{sd}$, we expect the data
below $\kappa_{sd}$ to display a stronger size dependence, and
hence to be less useful for the purpose of extrapolation. Although
due to the limitation on particle number, our study of the gap is
limited to $\kappa\gtrsim .8$, it is apparent that the energy gap
does not decline as the order parameter becomes exponentially
small. In addition to the above numerical evidence, there is an
extra reason to believe that for the bosonic non-abelian state the
gap cannot be caused by symmetry breaking, because there is a
totally symmetric ground state at all values of $\kappa$. 
To summarize, both the numerical results for the
order parameter and the energy gap  are consistent with the notion
of adiabatic continuity between small and large $\kappa$.

{\bf Quasiparticles~} Again, we first understand the quasiparticle
and quasihole in the large $\kappa$ limit. The most naive way to
produce a charge $\Delta Q=1$ state is to remove one boson from
one of the ground states, say the (02), as shown in Fig.1(d).
(Using the nomenclature of the electron, we regard the boson as
negatively charged.) As for all the abelian cases, the positively
charged defects cost zero energy. A closer inspection of Fig.1(d)
reveals that such a defect is consisted of two domain walls
(between the ground state in Fig.1(a) and 1(c)) placed at the
closest distance. Since a string of $1$'s does not cost any
additional energy,  it is possible to separate the two domain
walls, without injecting any extra charge, by converting an even
sequence of $0202..02$ into a string of $1$'s as shown in
Fig.1(e). The two domain walls at the
end of the string each carry a charge $\Delta Q=1/2$.
Now let us consider removing a boson from the (11) ground state.
By turning an odd string of $1$'s into $0202....20$ as sown in
Fig.1(f) two domain walls of charge $\Delta Q=1/2$ are produced. A
similar procedure can lead to $\Delta Q=-1/2$, nonzero energy,
anti-domain walls. Since there is adiabatic continuity
 we should be able to tune $\kappa$ down and let each
of these domain walls evolve into the fractional charged
quasiholes (delocalized in the $L_1$-direction) in the non-Abelian
Hall liquid. We note that the quasi-particle/quasi-hole charge of
$\Delta Q=\pm 1/2$, which follows naturally from our simple large
$\kappa$ analysis, indeed agrees with the prediction in
Ref.\cite{mooreread}.

Next we consider four quasiholes. In Fig.1(g) and (h) two distinct
four domain wall configurations (on top of the (02) ground state)
are presented. These patterns are differentiated by the fact that
in Fig.1(g)/(h) the middle string of $0,2$'s are in phase/out of
phase with those at the two ends. To transform panel (g) to (h)
one has to move the $1$ at the left end of the middle $0,2$ string
to the right end and shift the entire $0,2$ string to the right.
For far separated domain walls this move clearly involves many
particles and is very non-local. Comparing Fig.1(g) and (h) one
might regard the two middle domain walls as having different
positions (a shift of one lattice constant). However this
distinction will cease to exist if we let $\kappa$ decrease so
that further range hopping can take place. In that case each of
the domain wall discussed above will smear into wider and smoother
transition region between different ground states. The width of
the domain wall increases as $\kappa$ decreases. Although the tiny
difference in the positions of the domain wall becomes immaterial
as $\kappa$ decreases, the fact that in the case of
infinitely-separated domain walls no finite particle and local
moves can transform Fig.1(g) into (h) suggest that they will
remain distinct. This results in a two-fold degeneracy for four
quasihole state at fixed positions. Similarly for six quasiholes,
we have three disconnected string of $1$'s. In the gap between the
first and second, and second and third strings (of $1$'s) we can
again insert either (02) or (20) ground states. Thus for six
qusiholes at fixed positions, we have a degeneracy of 4.
Continuing this line of argument, we conclude that in the presence
of $2n$ quasiholes, we have a $2^{n-1}$ fold degeneracy. Similar
arguments can be carried out for the quasihole injection into the
other two ground states. This is consistent with the prediction of
the wavefunction analysis\cite{nayak}.

The above arguments lead us to an interesting observation. There
is an intricate difference between the two-domain-wall state in
the (02) and (20) background and the (11) background. Let us first
focus on the (11) background. Fig.1(i) and (j) represent two
patterns of domain wall configurations differing by one lattice
constant translation. Since we have removed a boson, which changes
the filling factor away from 1, the $U$ eigenvalues of these two
patterns are different. Hence they can not be mixed by further
range hopping when we decrease $\kappa$ in \Eq{H}. Now let us
consider the (02) background. It is simple to convince oneself
that due to the symmetry breaking and the periodic boundary
condition we can only shift the two domain walls by even multiple
lattice constants.  Consequently there are twice as many 
two-domain-wall states in the (11) background as in the (02) and (20)
background. Interestingly, 
the analysis of
Greiter {\it et al} shows that for fixed quasihole positions,
including all ground states, the two-quasihole-state is four-fold
rather than three-fold degenerate. We believe that the 
extra
degeneracy is due to the existence of twice as many two-domain wall
states that can be added to the (11) ground state.

In conclusion, we have shown that the bosonic non-abelian Hall
liquid is adiabatically connected to simple density patterns as
the small dimension of the torus decreases. Together with our
earlier result on the abelian Hall liquid\cite{seidel,hr,karl2}, a
powerful conclusion emerges. The paradigms of quantum number
fractionalization in one dimension (polyacetylene) and two
dimension (fractionally quantum Hall effect) are in fact
equivalent.

Acknowledgement: In the process of preparing this manuscript, we
become aware of a preprint by Bergholtz {\it et al}\cite{bergholtz}
on a closely related subject for the $\nu=1/2$ fermionic
pfaffian state.  DHL and AS are supported by DOE grant
DE-AC03-76SF00098.

\end{document}